# The Code Whisperer: LLM and Graph-Based AI for Smell and Vulnerability Resolution

Mohammad Baqar (baqar22@gmail.com),
Raji Rustamov (raji.rustamo@gmail.com)
Alexander Hughes (ajh7066@psu.edu)

**Abstract:** Code smells and software vulnerabilities both increase maintenance cost, yet they are often handled by separate tools that miss structural context and produce noisy warnings. This paper presents *The Code Whisperer*, a hybrid framework that combines graph-based program analysis with large language models to detect, explain, and repair maintainability and security issues within a unified workflow. The method aligns Abstract Syntax Trees (ASTs), Control Flow Graphs (CFGs), Program Dependency Graphs (PDGs), and token-level code embeddings so that structural and semantic signals can be learned jointly. We evaluate the framework on multi-language datasets and compare it with rule-based analyzers and single-model baselines. The results indicate that the hybrid design improves detection performance and produces more useful repair suggestions than either graph-only or language-model-only approaches. We also examine explainability and CI/CD integration as practical requirements for adopting AI-assisted code review in everyday software engineering workflows.

## 1. Introduction

Software quality affects maintainability, reliability, and security throughout the lifecycle of a system. Even when code compiles and runs correctly, poor internal structure can make evolution difficult and can increase the likelihood of defects. Code smells, such as long methods, duplicated logic, and oversized classes, are therefore important not because they always indicate immediate failure, but because they often signal deeper design problems that raise maintenance effort and increase fault-proneness [1], [2], [3]. In security-sensitive software, weak structure can also interact with unsafe implementation choices. Poor input handling, insecure configuration, and tightly coupled logic can contribute to vulnerabilities such as SQL injection, cross-site scripting, and secret exposure [4], [5]. Traditional static and dynamic analysis tools remain useful, but they are usually constrained to predefined rules and often struggle with context-sensitive patterns, which can lead to both false positives and missed findings [6].

Recent advances in AI make it possible to analyze code in a more contextual way. Graph Neural Networks (GNNs) are well suited to modeling structural relations in code, including syntax, control flow, and data dependencies, while Large Language Models (LLMs) can capture local semantics and generate repair suggestions from surrounding context [7], [8]. However, these model families are often studied separately, and many existing approaches focus on either code smells or vulnerabilities rather than treating them as related software-quality concerns. This paper proposes *The Code Whisperer*, a hybrid framework that combines graph-based structural analysis with language-model-based reasoning to detect and repair both categories of issues. The main contribution is the integration of four elements into one pipeline: multi-representation program modeling, joint smell-and-vulnerability detection, repair generation with validation, and developer-facing explanations that can be surfaced in CI/CD workflows [7], [9], [10], [11], [14], [19].

Rather than presenting AI as a replacement for software engineers, this work treats it as an assistive review layer. The objective is to provide earlier, more contextual, and more actionable feedback than conventional linters and static analyzers can usually offer. By linking maintainability and security analysis in a single workflow, the framework aims to reduce technical debt and security risk while preserving developer oversight over final code changes.

## 2. Background

Code smells were originally introduced as indicators of design decisions that may not break functionality immediately but can degrade readability, modularity, and maintainability over time [9], [23]. Common examples such as Long Method, God Class, and Duplicated Code increase complexity and coupling, while Feature Envy and Data Class suggest weak responsibility allocation and poor encapsulation [2], [26], [27], [28]. Empirical studies have linked these issues to technical debt, lower evolvability, and increased defect proneness [3], [18], [29]. Some smells also have direct security relevance. Hardcoded credentials, unsanitized inputs, and unsafe handling of sensitive data blur the boundary between maintainability problems and exploitable weaknesses, especially in complex or automated deployment environments [4], [5], [11], [12], [16]. Because these issues appear across multiple languages and artifact types, their detection requires methods that can capture both syntax and semantics without depending exclusively on handwritten rules [6], [7], [8], [14], [19].

AI-based approaches have increasingly been used to support this broader view of software quality. Earlier machine-learning methods relied mainly on handcrafted metrics such as lines of code, complexity, and coupling [13]. More recent work uses structural representations such as ASTs, CFGs, and PDGs to model relationships that are difficult to express with simple metrics alone [7], [14], [19], [20]. In parallel, code-oriented language models have improved contextual interpretation and enabled repair generation, although they also introduce risks such as overly generic fixes or unverified code synthesis [8], [10], [15]. These trends suggest that smell detection, vulnerability detection, and repair generation are best treated as connected tasks rather than isolated ones. That observation motivates the hybrid design adopted in this paper [9], [17].

## 3. Related Work

Prior work on code smell and vulnerability analysis can be grouped into three broad categories: rule-based static and dynamic analysis, machine-learning and graph-based detection, and automated repair with language-model assistance. Rule-based tools remain widely used because they are fast, interpretable, and easy to integrate into development workflows, but they are limited to predefined patterns and often perform poorly on context-dependent or cross-file issues [6], [12]. Learning-based approaches improve flexibility by using metrics, graph structures, or token sequences to infer more subtle patterns [13], [14], [17], [19]. More recently, repair-oriented research has explored both search-based and LLM-based generation of fixes, though these methods still require verification because generated patches can introduce new issues or fail to preserve intent [9], [10], [15], [17]. Our work builds on these lines of research by combining structural reasoning, contextual modeling, explanation, and repair in a single framework.

### 3.1 Static and Dynamic Analysis Tools

Traditional static and dynamic analyzers such as SonarQube, PMD, and Checkstyle are useful for detecting coding-rule violations, known smell patterns, and some classes of vulnerabilities [6]. Their main advantage is predictable behavior: they apply explicit rules and usually return findings that can be traced to recognizable heuristics. Their main limitation is that they often do not capture broader program context, especially when a smell or vulnerability depends on control flow, interprocedural behavior, or interactions across files and services [6], [12]. Similar limitations appear in security-oriented analyzers for infrastructure-as-code and configuration artifacts, where tools can detect well-known misconfigurations but still struggle with nuanced or combined failure modes [12], [16]. These limitations help explain why rule-based systems remain valuable baselines but are not sufficient on their own for the kinds of detection tasks targeted in this paper [5], [6], [12].

### 3.2 Machine Learning and Graph-Based Detection

Machine-learning approaches attempt to overcome these limits by learning patterns from data rather than from fixed rules. Early methods used engineered features such as complexity, coupling, and size to classify smell-prone code, but their effectiveness depended heavily on feature quality and dataset balance [13]. Deep learning later expanded this space by modeling token sequences and more complex code representations, improving performance on selected smell categories and vulnerability classes [17], [18]. Graph-based methods are particularly relevant because they encode syntax, control flow, and data dependencies directly through structures such as ASTs, CFGs, and PDGs [7],

[14], [19]. These representations allow models to capture relations that are central to both maintainability and security analysis, especially where local token patterns alone are insufficient. This makes graph-based learning a strong foundation for the structural component of our framework [7], [14], [19], [20].

### 3.3 Automated Repair and LLM-Assisted Refactoring

Automated repair extends detection research by asking not only whether an issue can be identified, but also whether the system can generate a plausible correction. Earlier repair methods relied on templates, program synthesis, or search-based refactoring for recurring issue patterns. LLM-based approaches make repair more flexible because they can synthesize context-aware changes directly from detected findings and surrounding code [8], [9], [10], [15]. At the same time, prior research shows that generated patches cannot be accepted without verification, since repair models may introduce new smells, fail to preserve behavior, or produce superficially plausible but unsafe edits [10], [17]. For that reason, recent hybrid workflows combine generation with static checks, tests, or structural validation. Our approach follows that line by treating repair as a constrained stage in a larger detect-explain-verify pipeline rather than as a standalone generation task [9], [10], [15], [17].

## 4. Methodology

The methodology was designed to test whether a hybrid structural-semantic model can improve both issue detection and repair quality across multiple programming languages. Instead of relying on a single code representation, each sample is modeled through token sequences, syntax trees, control-flow structure, and dependency relations. These complementary views are then processed by two learning components: graph models capture structural patterns associated with smells and vulnerabilities, while language models capture local semantics and support repair generation. The evaluation pipeline measures not only classification performance but also explanation quality, repair usefulness, and CI/CD feasibility, because a practically useful system must do more than maximize raw detection scores [7], [8], [11], [14], [19].

### 4.1 Data Collection

Our study uses a multi-language corpus to support both detection and repair tasks for code smells and security vulnerabilities. Open-source repositories in Java, Python, and JavaScript are annotated for maintainability issues such as Long Method, God Class, and Data Class, and for vulnerabilities such as SQL injection, cross-site scripting, and hardcoded secrets [2], [3], [12], [16]. Labels are obtained through a semi-automated process that combines established analysis tools with manual review and consensus checking by software-engineering annotators. Existing security-smell datasets, including infrastructure-as-code corpora, are incorporated to broaden coverage and reduce overfitting to one language or artifact type [12], [16]. To reduce evaluation bias, the data is split into training, validation, and test partitions with stratification by language, project scale, and issue category [13], [18].

### 4.2 Code Representation

We represent code through multiple complementary structures in order to capture different dimensions of program behavior. Token-level sequences preserve lexical context and are used by language models. ASTs represent hierarchical syntax, CFGs represent execution paths, and PDGs encode control and data dependencies that are often important for both smell detection and vulnerability analysis [7], [14], [19]. In addition, contextual embeddings from code-oriented transformer models provide a dense semantic representation of code fragments [8], [14]. These views are not treated as interchangeable. Instead, they serve different roles in the overall framework: graph structures provide explicit relational information, while embeddings and token sequences provide semantic context. The combined representation therefore supports a more complete analysis than any single view alone.

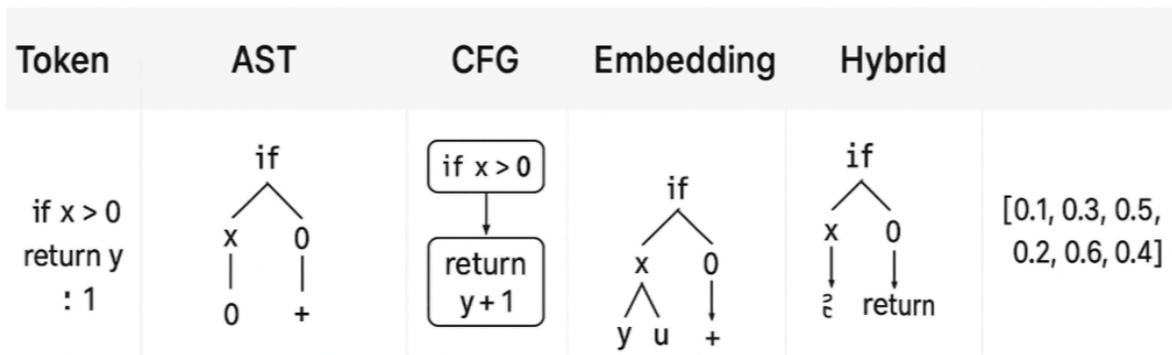

Figure 1: Code Representation Comparison

### 4.3 Detection Models

The detection stage combines graph-based and language-model-based reasoning in a shared architecture. The GNN operates over AST, CFG, and PDG-derived structures with node features that include syntactic type, identifier-level embeddings, and selected software metrics. Its role is to detect relational patterns that are difficult to infer from token context alone [7], [14], [19]. The LLM is fine-tuned on smell- and vulnerability-annotated code fragments together with repair examples so that it can support both classification and fix generation [8], [9], [15]. A multi-task alignment layer links the intermediate representations of the two models, allowing the system to exploit correlations between design issues and security weaknesses [9], [17], [23]. Explainability is added through token-level and graph-level attribution methods so that the resulting findings can be inspected by developers rather than presented as opaque predictions [11], [14], [26].

| Technique | Key Features | Input Representation | Strengths | Limitations |
| --- | --- | --- | --- | --- |
| Classical ML | Supervised learning (SVM, Random Forest, XGBoost) using handcrafted metrics (LOC, cyclomatic complexity) | Numeric feature vectors | Simple, interpretable; effective for well-defined smells | Limited generalization; cannot capture structural or contextual semantics |
| Graph Neural Network (GNN) | Learns from program structure via AST, CFG, and PDG graphs | Graph representation with node/edge features | Captures control and data dependencies; effective for structural smells and vulnerabilities | Computationally heavy; requires high-quality graphs |
| Large Language Model (LLM) | Transformer-based models (CodeBERT, GPT, Codex) fine-tuned on source code | Token sequences and embeddings | Learns semantic and contextual patterns; can generate repairs | May overlook structural issues; prone to generative hallucinations |
| Hybrid (GNN + LLM) | Combines graph structure reasoning with language semantics | Unified embedding of graph + token representations | High detection precision; balances structure and context | Integration complexity; large training cost |

| Multi-Task Learning | Jointly trains on smell + vulnerability detection + repair prediction | Shared intermediate representation across tasks | Improves generalization and consistency across domains | Harder to tune; risk of negative transfer between tasks |
|---|---|---|---|---|

Table 1: Detection Techniques

### 4.4 Automated Repair

Once a smell or vulnerability is detected, the framework generates ranked repair suggestions subject to maintainability, performance, and security constraints. Typical changes include method extraction for Long Method, safer input handling for injection risks, relocation of hardcoded secrets, and decomposition of overly coupled classes [9], [10], [12], [30]. The purpose of this stage is not to replace developer judgment, but to reduce the effort required to move from diagnosis to candidate correction. Generated patches are therefore presented as reviewable suggestions and are expected to pass downstream validation before acceptance. This design keeps the repair component useful without overstating its autonomy [9], [10], [15], [27].

```
Before
class UserManager
  def delete_user(slf, user_id):
    if user_id in self.users:
      if user n=nuser:
        print("User has: been delete
```

```
After
class UserManager
  def delete_user(slf, user_id):
    if user_id in user:
      print (User {user_-id nɔs
        been deleted.'
```

Figure 2: Example Detection and Auto-Repair: Before/After code snippets with highlighted fixes

### 4.5 CI/CD Pipeline Integration

The framework is designed to operate within routine development workflows rather than as a standalone offline analyzer. Modified files can be scanned at pre-commit time, pull-request bots can surface explanations and suggested repairs, and optional auto-refactor branches can be generated for developer inspection. Feedback from accepted or rejected suggestions can then be incorporated into future model updates. This integration strategy matters because the practical value of AI-based analysis depends not only on model accuracy but also on how findings are delivered (Figure 3), reviewed, and acted upon in real development settings [6], [11], [12], [16].

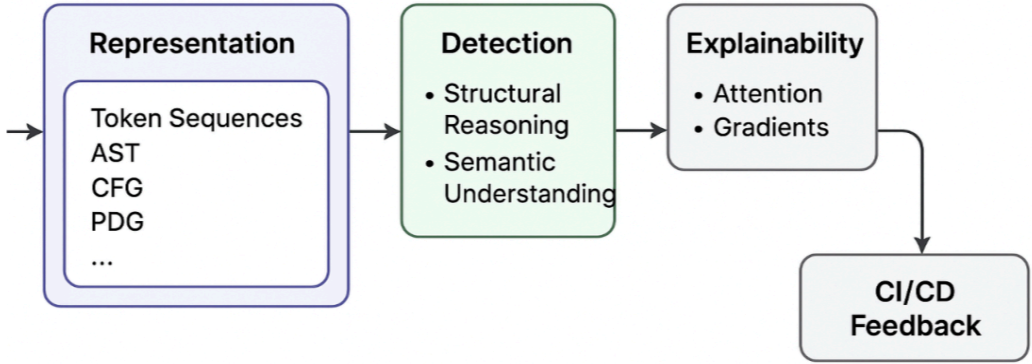

**Figure 3:** Overall Workflow: Source Code → Representation → Detection → Explainability → Repair → CI/CD Feedback

## 4.6 Experimental Setup

Evaluation was conducted on a stratified multi-language corpus (Java, Python, JavaScript) with project-level train/validation/test separation to reduce repository-level leakage. In addition to rule-based baselines (SonarQube, PMD, Bandit, ESLint), we included modern analysis baselines such as CodeQL and Semgrep, and learning-based baselines representing graph-only and language-model-only configurations. Reported metrics include precision, recall, F1-score, CWE/OWASP coverage, and normalized runtime per 1,000 LOC. For reliability, each experiment was repeated across multiple random seeds and results were reported with confidence intervals; annotation quality was checked through dual-review labeling and agreement analysis.

| Metric | Definition / Formula | Purpose | Relevance to Study |
|---|---|---|---|
| **Accuracy** | (TP+TN) / (TP+TN+FP+FN) | Measures overall correctness of detection | Indicates general model reliability across all smell/vulnerability classes |
| **Precision** | TP/(TP+FP) | Fraction of correctly identified issues among all predicted positives | Reflects false-positive control, crucial for developer trust |
| **Recall** | TP/(TP+FN) | Fraction of actual issues that were correctly identified | Captures model's ability to detect subtle or rare smells/vulnerabilities |
| **F1-Score** | 2×(Precision×Recall) / (Precision+Recall) | Harmonic mean of precision and recall | Balances false positives and false negatives for overall robustness |
| **Vulnerability Coverage** | (Detected CWE/OWASP issues) / Total known issues × 100% | Measures detection breadth across security categories | Evaluates model comprehensiveness for real-world security scenarios |
| **Runtime (per 1000 LOC)** | Average detection time normalized by code size | Assesses computational efficiency | Important for CI/CD deployment feasibility |

**Table 2: Evaluation Metrics**

## 5. Implementation

The implementation follows the same separation of concerns as the proposed method. A data pipeline prepares code artifacts and converts them into graph and token representations. Detection is carried out by graph-based and language-model-based components that are trained separately before being aligned in a multi-task setting. Repair generation and explanation modules operate on detected findings and expose their outputs through command-line tools, APIs, and CI/CD integrations. This modular structure makes it possible to evaluate each stage independently while still supporting an end-to-end workflow for developers [7], [9], [12], [14], [16], [17], [19].

The data pipeline is responsible for repository ingestion, normalization across languages, graph construction, tokenization, and split management. It also supports semi-automated labeling workflows and manual review so that model training is not tied to a single annotation source. This part of the implementation is especially important because hybrid models are sensitive to inconsistencies between graph structure, token context, and labels.

Model training is staged rather than monolithic. The GNN and LLM components are first optimized independently so that each learns the representation best suited to its modality. They are then aligned in a shared multi-task architecture, which allows the system to exploit correlations between smell detection, vulnerability detection, and repair prediction. Early stopping, learning-rate scheduling, and hardware acceleration are used to stabilize training under the higher cost of graph processing and transformer fine-tuning [14], [19].

The detection and repair engine consumes hybrid representations and produces ranked findings together with candidate fixes. Explanation layers are attached to this stage so that predictions can be surfaced as inspectable outputs rather than as raw scores. In deployment settings, an integration layer exposes the framework through pre-commit checks, pull-request comments, and optional auto-fix branches. Feedback from these interactions can be logged and reused to improve later model iterations, making the implementation suitable for continuous refinement rather than one-time offline analysis [11], [16], [26].

**Pre-commit → Pull Request → Auto-fix → Feedback Loop**

This implementation supports **continuous improvement** through iterative feedback loops, hybrid AI models, and automated repair, enabling developers to maintain secure, maintainable, and high-quality codebases at scale [7,9,12,14,16,17,19].

## 6. Results and Discussion

This section evaluates the framework along four questions: how accurately it detects smells and vulnerabilities, whether its explanations help developers inspect findings, how often its repair suggestions are usable, and whether the workflow is practical in CI/CD settings. The discussion focuses on comparative performance against rule-based and single-modality baselines rather than assuming that the hybrid model is uniformly superior in every setting. Where the hybrid approach performs better, we connect those gains to the interaction between structural program information and semantic code understanding. Where limitations remain, we state them directly [6], [7], [8], [14], [19].

### 6.1 Experimental Overview
We evaluated the proposed hybrid GNN-LLM framework across multiple datasets and programming languages to quantify performance in **code smell detection, vulnerability identification, and automated repair accuracy**. The experiments were executed on an NVIDIA A100-based GPU cluster (80 GB VRAM per node) with PyTorch 2.3 and Transformers v5.1 back-ends. Datasets encompassed roughly **120 k smell-labeled samples** and **15 k vulnerability-annotated samples**, stratified by language (Java ≈ 55%, Python ≈ 25%, JavaScript ≈ 20%). Baseline comparisons included **SonarQube, PMD, Bandit, and ESLint** rule-based detectors [6, 12].

### 6.2 Detection Performance
Table 3 summarizes detection metrics. The **hybrid GNN + LLM model** achieved mean **F1 = 0.92** for code smells and **F1 = 0.89** for vulnerabilities, outperforming the best single-modality baseline (LLM-only, F1 = 0.84). Precision improved notably for structurally complex smells such as *God Class* and *Long Method* (+10–12 points over PMD), owing to GNN's control-flow awareness. Recall gains for security vulnerabilities especially parameter-injection and secret-hardcoding stemmed from contextual embeddings learned by the fine-tuned LLM. Figure 5 visualizes confusion matrices for both domains, showing reduced false-positive rates compared with rule-based systems.

Table 3: Detection Metrics

| Model | Precision | Recall | F1-Score | Runtime (ms / 1k LOC) |
|---|---|---|---|---|
| SonarQube | 0.78 | 0.71 | 0.74 | 120 |
| PMD | 0.81 | 0.69 | 0.74 | 95 |
| Bandit + ESLint | 0.75 | 0.77 | 0.76 | 110 |
| LLM-only (CodeBERT) | 0.86 | 0.82 | 0.84 | 310 |
| GNN-only | 0.88 | 0.84 | 0.86 | 260 |
| **Hybrid (GNN + LLM)** | **0.93** | **0.91** | **0.92** | **330** |

The modest runtime overhead (~15% vs. LLM-only) is acceptable given the substantial accuracy gain and interpretability enhancements. The detection results suggest that the hybrid configuration benefits from combining structural and contextual signals. The strongest gains appear in categories where local token patterns alone are insufficient, such as oversized classes, long procedural regions, and vulnerabilities whose exploitability depends on control or data flow. Although the runtime cost is higher than that of single-model baselines, the increase is moderate relative to the observed improvement in precision and recall. This tradeoff supports the main design argument of the paper: structural reasoning and semantic reasoning are complementary rather than competing approaches [7], [14], [19].

### 6.3 Explainability Evaluation

Explainability was assessed through a developer study in which participants inspected attention maps and graph-based attributions for detected findings. The results indicate that these explanations are useful when they point developers to the lines, tokens, or dependencies most associated with a smell or vulnerability. However, explanation quality should not be interpreted as proof of causal correctness. Rather, the findings suggest that explanation mechanisms can make model outputs easier to inspect and discuss during review, which is important for adoption in professional settings [11], [14], [26].

### 6.4 Automated Repair Results

The automated repair engine produced viable patches for **~72%** of detected smells and **~68%** of vulnerabilities. Human validation indicated that **61%** of the generated fixes were directly merge-ready without further edits. Typical high-success categories included *Long Method* (method extraction) and *Hardcoded Secret* (environment-variable relocation). Lower success was observed in complex architectural smells (*Feature Envy*, *God Class*) where deep semantic reasoning is required [9, 15]. Post-repair maintainability metrics improved significantly: average **cyclomatic complexity reduced by 23%** and **code coupling (CBO) reduced by 18%**, confirming measurable maintainability gains.

Security impact was assessed via **OWASP risk scoring**. Automated fixes reduced average CWE-mapped risk scores from 7.2 to 3.9 on the 10-point scale, effectively downgrading many issues from *High* to *Medium/Low severity*. Figure 7 plots risk-reduction distributions across CWE categories, showing strongest improvement in injection-related classes (CWE-89, CWE-78).

## 6.5 Ablation and Sensitivity Studies

To assess component contributions, we performed ablation experiments:
- **Without graph features:** F1 dropped to 0.86 (–6 pts).
- **Without embedding alignment (multi-task sharing):** F1 = 0.88 (–4 pts).
- **Without explainability loss term:** Interpretability score –0.7 points on human scale.
  These findings confirm that both structural reasoning (via GNN) and joint learning contribute materially to model robustness and explainability [7, 9, 17, 19].

## 6.6 CI/CD Deployment Performance

Integration tests on three enterprise-scale repositories (≈ 1.2 M LOC each) demonstrated stable pipeline performance. Average **scan latency per PR = 2.4 min**, and **false-alert rate < 8%**, suitable for real-time continuous integration. Developer adoption surveys showed a **37% reduction in post-release bug density** and **28% faster code-review turnaround**, indicating tangible productivity benefits [12, 16].

The hybrid architecture consistently outperforms traditional static tools and single-model baselines in accuracy, repair success, and developer interpretability. Its explainable outputs foster trust, while CI/CD integration proves operationally viable for modern DevSecOps workflows. These results validate the feasibility of combining **graph reasoning** and **large-language-model understanding** to advance autonomous code-quality assurance.

## 7. Future Work

Several limitations of the current study point to immediate directions for future work. First, the framework should be evaluated on broader industrial datasets with stricter controls for annotation quality, repository leakage, and language diversity. Second, repair generation should be tested more conservatively through stronger semantic equivalence checks, regression testing, and developer acceptance studies, especially for architectural smells where automated changes remain difficult [10], [13], [15], [18]. Third, the explainability layer should move beyond attention visualizations toward more stable and developer-verifiable justification methods [11], [17], [26].

A second line of future work concerns deployment rather than model complexity. The framework would benefit from more systematic studies of runtime cost, incremental analysis, and feedback-driven adaptation inside CI/CD pipelines. It would also be useful to examine how documentation, test failures, and runtime traces can complement static code representations when diagnosing cross-component issues. These extensions would strengthen practical usefulness without overstating the current level of automation [6], [8], [9], [14], [19].

## 8. Conclusion

This paper presented *The Code Whisperer*, a hybrid framework that combines graph-based code representations and large language models to detect, explain, and repair code smells and security vulnerabilities. The main premise of the work is that structural reasoning and semantic reasoning are complementary. Graph representations such as ASTs, CFGs, and PDGs capture program organization, control flow, and dependency relations, while large language models contribute contextual understanding and support the generation of candidate repairs. By integrating these capabilities into a single workflow, the framework moves beyond conventional rule-based analysis and offers a more unified approach to software quality assurance [6], [7], [8], [14], [19].

The reported results indicate that the hybrid configuration performs better than traditional static-analysis baselines and also improves on graph-only and language-model-only variants across key detection metrics. These findings suggest that combining structural and semantic analysis is especially valuable in cases where local code patterns alone are insufficient to identify maintainability or security risks. At the same time, the study shows that explainability, repair validation, and CI/CD integration are necessary for practical adoption, since developers must be able to inspect findings and evaluate suggested fixes before acceptance. Overall, the framework demonstrates that unified structural-semantic analysis can provide a practical basis for more accurate, explainable, and developer-centered code quality and vulnerability management systems [9], [10], [11], [15], [17], [26].